\documentclass[usenatbib]{mn2e}    

\usepackage{graphicx}
\usepackage{float}
\usepackage{fancyhdr}
\usepackage{natbib}
\usepackage{amsmath}
\usepackage{amssymb}
\usepackage[usenames,dvips]{color}
\usepackage{appendix}
\usepackage{subfigure}

\newcommand{\pth}{P_\text{thermal}}
\newcommand{\pram}{P_\text{ram}}
\newcommand{\msun}{M_\odot}

\defcitealias{Crain_et_al_2009}{C09}
\defcitealias{Mulchaey_Jeltema_2010}{MJ10}
\defcitealias{David_et_al_2006}{D06}
\defcitealias{Sun_et_al_2007}{S07}
\defcitealias{Jeltema_et_al_2008}{JMB08}

\title[Confinement pressure]{The competition between confinement and ram pressure and its implications for galaxies in groups and clusters}
\author[Y.M. Bah\'{e} et al.]{Yannick~M.~Bah\'{e}\thanks{ybahe@ast.cam.ac.uk}$^1$, Ian G.~McCarthy$^{3,2,1}$, Robert A. Crain$^4$ and Tom Theuns$^{5,6}$\\
$^1$ Institute of Astronomy, University of Cambridge, Madingley Road, Cambridge CB3 0HA, United Kingdom\\
$^2$ Kavli Institute for Cosmology Cambridge, University of Cambridge, Madingley Road, Cambridge CB3 0HA, United Kingdom\\
$^3$ School of Physics and Astronomy, University of Birmingham, Edgbaston, Birmingham B15 2TT, United Kingdom\\
$^4$ Leiden Observatory, Leiden University, P. O. Box 9513, 2300 RA Leiden, the Netherlands\\
$^5$ Institute for Computational Cosmology, Department of Physics, University of Durham, South Road, Durham DH1 3LE, United Kingdom\\
$^6$ Department of Physics, University of Antwerp, Campus Groenenborger, Groenenborgerlaan 171, B-2020 Antwerp, Belgium
}
\begin{document}
\label{firstpage}
\maketitle

\begin{abstract}
Ram pressure stripping of the hot gas that surrounds normal galaxies as they fall into groups and clusters (also referred to as `strangulation' or `starvation') is generally thought to shut down star formation on a time scale of a few Gyr.  However, it has recently been suggested, on the basis of X-ray-optical scaling relations of galaxies in the field and the group/cluster environment, that confinement pressure by the intra-cluster medium can actually lead to an {\it increase} in the mass of hot gas surrounding these galaxies.  We investigate the competition between pressure confinement and ram pressure stripping for satellite galaxies in orbit about galaxy groups and clusters using simple analytic models and detailed cosmological hydrodynamic simulations.  It is found that, independent of host mass, ram pressure is generally dominant over confinement pressure --- only $\sim 16\%$ of galaxies find themselves in the reverse situation.  Furthermore, these galaxies have, on average, \emph{less} hot gas than ram-pressure dominated ones, contrary to simple expectations.  This is explained by the fact that the small number of galaxies which are confinement dominated are typically at first or second apocentre and have therefore already been maximally affected by ram pressure stripping around first pericentre. Our results are shown to be insensitive to host halo mass; we argue that the same is true for uncertain sub-grid processes, such as feedback.
\end{abstract}

\begin{keywords}
galaxies: clusters: general --- galaxies: evolution --- galaxies: haloes --- galaxies: interactions --- galaxies: intergalactic medium --- galaxies: ISM
\end{keywords}

\section{Introduction}
\label{sec:introduction}

In the current paradigm of galaxy formation, haloes of hot gas are predicted to be a common feature around galaxies with masses similar to (or larger than) the Milky Way (\citealt{White_Rees_1978}; \citealt{White_Frenk_1991}). With a temperature of $\sim 10^6$ K, they are diffuse sources of soft X-ray emission. There is now a large number of observational detections of these ``X-ray coronae'', both around normal elliptical (e.g.; \citealt{Forman_et_al_1985}; \citealt{Kim_et_al_1992}; \citealt{Osullivan_et_al_2001}; \citealt{David_et_al_2006}; \citealt{Sun_et_al_2007}; \citealt{Jeltema_et_al_2008}; \citealt{Sun_et_al_2009}) and recently also around both normal and star bursting disc galaxies (\citealt{Strickland_et_al_2004}; \citealt{Wang_2005}; \citealt{Tuellmann_et_al_2006}; \citealt{Li_et_al_2006}; \citealt{Li_et_al_2007}; \citealt{Sun_et_al_2009}; \citealt{Owen_Warwick_2009}; \citealt{Anderson_Bregman_2011}; \citealt{Li_Wang_2012}). Observed at first only in galaxies in the field and poor environments \citep{Forman_et_al_1985}, improved observing facilities like the \emph{Chandra} and \emph{XMM-Newton} telescopes have established the presence of hot gas haloes also in galaxies in groups \citep{Jeltema_et_al_2008} and even clusters, where \citet{Sun_et_al_2007} found them to be as common as $> 60\%$ in $L_{K_S} > 2L_*$ galaxies.     

Studying these X-ray coronae promises to enhance our understanding of galaxy formation and evolution, as one central prediction from the theoretical models of \citet{White_Rees_1978} and \citet{White_Frenk_1991} is that the hot gas cools and replenishes the cold gas reservoir which is responsible for fuelling star formation. One complication, however, is that not only gas which is cooling and thus \emph{inflowing} into the galaxy centre can emit X-rays, but so can outflowing hot gas driven by e.g.~supernovae or active galactic nuclei (AGN). There is as yet no clear consensus on which of these two processes dominates the X-ray emission: while it has long been assumed to be the latter (see, e.g., \citealt{Read_Ponman_1998}; \citealt{Mathews_Brighenti_1998}), recent work by \citet[see also \citealt{Crain_et_al_2010b}]{Crain_et_al_2010a}, using hydrodynamical simulations which reproduce the observed X-ray-optical scaling relations of normal disc galaxies, indicates that the bulk of the X-ray emission comes from the cooling of the hot, quasi-hydrostatic corona.  Direct confirmation of the nature of the X-ray emitting gas will be provided in the future by deep, high-resolution spectra from observatories such as \emph{IXO} or \emph{NeXT/ASTRO-H}.

Of course, galaxies are not just shaped by internal processes like gas cooling and outflows, they are also influenced by their local environment. For example, the tidal stress induced by a group or cluster on an infalling galaxy can lead to stripping or even total disruption which may account, at least partially, for the well-known morphology-density relation of galaxies in denser environments being preferentially of early type (e.g., \citealt{Moore_et_al_1996}). A second important influence is the interaction of the intragroup/-cluster medium (ICM) with the galactic gas: as the galaxy moves through the ICM, its gas experiences a drag or ram pressure, which can lead to stripping of both the cold, central gas disc (e.g., \citealt{Gunn_Gott_1972,Abadi_et_al_1999}) and the hot, extended gaseous halo (e.g., \citealt{Larson_et_al_1980,Balogh_et_al_2000,McCarthy_et_al_2008}).  The latter process removes the possibility for gas to cool and replenish the cold disc, leading to star formation fading away over a period of several Gyr and is therefore commonly called `starvation' or `strangulation'.  This process is widely believed to be (at least partially) responsible for the the observed relation between galaxy colour and environment (e.g., \citealt{Weinmann_et_al_2006,Font_et_al_2008}). For X-ray observations, this implies less massive hot gas haloes around galaxies in groups and clusters, and therefore lower X-ray luminosities, compared to galaxies of similar mass in the field.  

A different effect of the ICM on the hot gaseous coronae of galaxies has recently been proposed by \citet[MJ10]{Mulchaey_Jeltema_2010}. Motivated by observations which appear to indicate a relative \emph{excess} of X-ray emission by galaxies in groups and clusters, compared to galaxies of the same K-band luminosity in the field, these authors suggested that the very hot ($T_\text{ICM} \sim 10^8$ K) ICM could exert pressure on cluster galaxies and their relatively cool gas halos with $T_\text{halo} \sim 10^6$ K. This `confinement pressure' could prevent the outflows driven by supernovae, AGN or massive stars from leaving the galaxy and therefore potentially result in an \emph{increase} in the hot gas density in galaxies orbiting within groups and clusters compared to those of similar (stellar) mass in the field. 

In this paper, we aim to investigate the relative importance of ram pressure stripping and confinement pressure on group and cluster galaxies using the \textsc{Galaxies-Intergalactic Medium Interaction Calculation} (\textsc{gimic}; \citealt[C09]{Crain_et_al_2009}), a set of five high-resolution cosmological hydrodynamical re-simulations of the formation and evolution of galaxies in a wide range of large-scale environments taken from the Millennium Simulation \citep{Springel_et_al_2005}. Analysing simulated galaxies, rather than real ones, brings the key advantage that the physical quantities responsible for both ram and confinement pressure, particularly the mass-weighted temperature and galaxy 3D velocity, are easily available. A second benefit is the possibility to trace the evolution of individual galaxies over time, which, as we demonstrate in Section \ref{sec:results}, is key to understanding the potential effect of confinement pressure. Furthermore, a particular advantage of using \textsc{gimic} is that its field galaxies have already been shown to have properties in good agreement with observational data, such as scaling of X-ray luminosity with K-band luminosity, star-formation rate and disc rotation velocity \citep{Crain_et_al_2010a} and those of stellar spheroids around Milky Way mass disc galaxies (\citealt{Font_et_al_2011}; \citealt{McCarthy_et_al_2012}). Analysing a realistic simulation will thus allow us to make meaningful conclusions concerning galaxies in the real Universe. 

This paper is structured as follows: In Section 2 we describe the cosmological simulations and our data extraction method, and present our results from both \textsc{gimic} and a simple analytic model in Section 3. Our findings are summarised and discussed in Section 4. A flat $\Lambda$CDM cosmology with Hubble parameter $h =$ H$_{0}/(100\,{\rm km}\,{\rm s}^{-1}{\rm Mpc}^{-1}) = 0.73$, dark energy density parameter $\Omega_\Lambda = 0.75$ (dark energy equation of state parameter $w=-1$), and matter density parameter $\Omega_{\rm M} = 0.25$ is used throughout this paper.

\section{Simulations and analysis}
\label{sec:simulations}

\subsection{Simulations and sample selection}
We extract galaxy groups and clusters from the \textsc{Galaxies-Intergalactic Medium Interaction Calculation} suite of simulations (\textsc{gimic}; \citetalias{Crain_et_al_2009}), a set of five high-resolution (a baryon mass resolution of $m_\text{gas} \sim 1.16 \times 10^7 h^{-1} \msun$ with a gravitational softening that is 1 $h^{-1}$ kpc in physical space at $z \le 3$ and is fixed in comoving space at higher redshifts) re-simulations of nearly spherical regions of varying mean density extracted from the Millennium Simulation \citep{Springel_et_al_2005}.  The simulations were carried out with the TreePM-SPH code \textsc{GADGET-3} (last described in \citealt{Springel_2005}) and include prescriptions for star formation \citep{Schaye_DallaVecchia_2008}, metal-dependent radiative cooling \citep{Wiersma_et_al_2009a}, feedback and mass transport by Type Ia and Type II supernovae \citep{DallaVecchia_Schaye_2008}, as well as stellar evolution and chemodynamics \citep{Wiersma_et_al_2009b}.  The reader is referred to \citetalias{Crain_et_al_2009} (see also \citealt{Schaye_et_al_2010,Font_et_al_2011}) for a detailed description of the simulations. 

As we are interested in groups and clusters, we use the two highest density simulations, `$+1\sigma$' and `$+2\sigma$' (see \citetalias{Crain_et_al_2009}). Groups and clusters of galaxies were identified at redshift $z$ = 0 using a standard Friends-of-Friends (FoF) algorithm with linking length $b$ = 0.2 times the mean inter-particle separation.  We select all FoF groups with $M_{200} > 10^{13.0} \msun$, where $M_{200}$ is the mass within a spherical region of radius $r_{200}$, centered on the most-bound particle, in which the mean density is 200 times the critical density of the universe.  There are 69 systems in total in the $+1\sigma$ and $+2\sigma$ \textsc{gimic} regions, with masses in the range $13.0 \leq \log_{10} (M_{200}/M_\odot) < 15.1$. We explore below whether our results depend on total group/cluster mass, $M_{200}$.

Within the simulated groups and clusters, bound substructures are identified using the \textsc{subfind} algorithm of \citet{Dolag_et_al_2009}, that extends the standard implementation of \citet{Springel_et_al_2001} by including baryonic particles in the identification of self-bound substructures.  `Galaxies' are identified as self-gravitating substructures with total stellar mass of $M_* > 10^9 \msun$ (i.e., similar to the mass of galaxies typically identifiable in present observations of local groups and clusters). We exclude the central dominant galaxy (the `BCG') from our analysis, as we are interested in the competition between ram pressure stripping and confinement pressure on orbiting galaxies. Also excluded are `galaxy' identifications at $r < 0.02\, r_{200}$ which we found to be associated with transient substructure in the BCG rather than being independent objects. In the 69 groups and clusters we have selected there are 1447 galaxies that meet our selection criteria, with stellar masses in the range $9.0 \leq \log_{10} (M_*/M_\odot) < 12.3$. 

\subsection{Analysis}
For each of these galaxies we compute the ratio of confinement pressure to ram pressure at the position of the galaxy as follows.  The ram pressure is given by $\pram = \rho v^2$, where $\rho$ is the density of the surrounding intracluster or -group medium and $v$ the velocity of the galaxy relative to the ICM.  The confinement pressure is simply the thermal pressure exerted by the hot ICM and is thus given by the ideal gas law $\pth = nk_{\rm B}T$ where $n$ is the intracluster particle number density, $k_{\rm B}$ Boltzmann's constant and $T$ the temperature of the ICM surrounding the galaxy. Combining these two, we obtain 
\begin{equation}
\alpha = \frac{\pth}{\pram} = \frac{k_\mathrm{B}T}{\mu m_\mathrm{p} v^2}
\label{eq:prat}
\end{equation}
where $\mu$ is the mean molecular weight of the ICM and $m_{\rm p}$ the proton mass.  We compute the galaxy velocities $v$ relative to the mass-weighted average velocity of the group or cluster particles which implicitly assumes the ICM is at rest in the cluster centre of mass frame; for simplicity we assume a constant value of $\mu = 0.58$ throughout. For each group and cluster, we compute a spherically averaged radial temperature profile by binning the hot gas ($T > 10^5$ K) particles in bins of radial width $\Delta_r = 0.1\, r_{200}$. The ICM temperature at the position of each galaxy is then determined by linear interpolation. We have verified that using locally-determined ICM temperatures computed as the mass-weighted mean of all hot gas particles within a radial range $20\, \text{kpc} < r < 50\, \text{kpc}$ around the galaxy centre produces very similar results.

Note that the ratio $\alpha$ between confinement and ram pressure in equation \ref{eq:prat} depends only on the galaxy velocity $v$ and the ICM temperature $T$, it does {\it not} depend on the density distribution of the ICM or the structure of the galaxy itself. The velocity, in turn, depends only on the depth of the potential well of the group/cluster, which is dominated by dark matter, and to first order this is also true of the temperature of the ICM (e.g., \citealt{Voit_et_al_2002,Hansen_et_al_2011}).  Therefore, we expect our results to be insensitive to uncertain sub-grid processes such as star formation and feedback.  Furthermore, since the dark matter mass distribution is approximately self-similar, we also expect our results to be approximately independent of total group/cluster mass or redshift. We explicitly verify this below. 

\section{Results}
\label{sec:results}
\subsection{Analytic expectations}
\label{sec:analyticexpectations}
Before proceeding to an analysis of the cosmological hydrodynamic simulations, we can gain some insight by considering a simple spherical analytic cluster in which the cluster galaxies orbit with the typical velocity $v_{\rm circ}(r) = [G M_{\rm tot}(< r)/r]^{1/2}$ and the hot gas is in hydrostatic equilibrium within the cluster potential well.

Re-writing the equation for hydrostatic equilibrium,

\begin{equation}
\label{eq:hse}
\frac{\text{d}P}{\text{d}r} = -\frac{G M_{\rm tot}(< r) \rho}{r^2}
\end{equation}
as 
\begin{equation}
\frac{\text{d} \ln P}{\text{d} \ln r} = -v_{\rm circ}(r)^2\, \frac{\rho}{P},
\end{equation}
and using the ideal gas law 
\begin{equation}
P = \frac{\rho}{\mu m_\mathrm{p}}k_\mathrm{B}T
\end{equation}
with mean molecular weight $\mu$ and proton mass $m_{\rm p}$, we obtain
\begin{equation}
\frac{\text{d} \ln P}{\text{d} \ln r} = -v_\text{circ}(r)^2\, \frac{\mu m_\mathrm{p}}{k_\mathrm{B} T} = -\frac{1}{\alpha}
\end{equation}
where the last equality is from equation \eqref{eq:prat} above, assuming galaxies move at the circular velocity $v_\text{circ} (r)$. The typical ratio between thermal and ram pressure is therefore directly related to the slope of the logarithmic pressure profile of the host group or cluster. 

Furthermore, if the hot gas density distribution is assumed to trace that of the dark matter and both are described by a spherically symmetric power-law of the form
\begin{equation}
\rho \propto r^{-\beta}
\end{equation} 
then the mass enclosed within a radius $r$ follows
\begin{equation}
M(<r) = \int_0^r 4\pi r^2 \rho\, \text{d}r \propto r^{3-\beta}.
\end{equation}
From equation \eqref{eq:hse}, we then obtain
\begin{equation}
\frac{\text{d}P}{\text{d}r} = -\frac{GM(<r)}{r^2}\rho \propto -\frac{r^{3-\beta} r^{-\beta}}{r^2} 
\end{equation}
so that
\begin{equation}
P \propto r^{2-2\beta}
\end{equation}
and therefore from equation (5)
\begin{equation}
\label{eq:analyticfinal}
\alpha = - \left(\frac{\text{d} \ln P}{\text{d} \ln r}\right)^{-1} = \frac{1}{2-2\beta}.
\end{equation}

For the case of an isothermal sphere with $\beta = 2$, the ratio of confinement pressure to ram pressure is $\alpha$ = 1/2. For a more realistic NFW profile, in which the effective exponent varies between $\beta = 1$ in the innermost regions and $\beta = 3$ at large distances from the centre, $\alpha$ should vary as well: Equation \eqref{eq:analyticfinal} predicts asymptotic behaviour of $\alpha \rightarrow \infty$ for $r \rightarrow 0$, and $\alpha \rightarrow 0.25$ for $r \rightarrow \infty$, with $\alpha = 0.5$ at the cluster scale radius ($\sim 0.20-0.25\, r_{200}$).  

These analytic calculations therefore suggest that confinement pressure will generally be subordinate to ram pressure except in the very inner regions of galaxy groups and clusters.  However, there are several potentially important caveats to the above argument.  First, the assumption that galaxies move at a velocity of \emph{exactly} $v_{\text{circ}}(r)\,$ is clearly not correct and dispersion in velocity will cause scatter in the pressure ratio $\alpha$ which might lead to confinement pressure being important beyond the very centre.  Secondly, deviations from spherical symmetry and hydrostatic and virial equilibrium may be relevant.  Finally, the gas distribution will not follow that of the dark matter precisely and this will have an effect (albeit a small one) on the temperature of the ICM.

For an as realistic as possible answer to the question of whether confinement or ram pressure stripping is the more important influence of the ICM on galaxies, we therefore need to take these complicating factors into account as well. This requires use of a detailed hydrodynamic simulation of groups and clusters, such as \textsc{gimic}.  

\subsection{Relative importance of confinement and ram pressure in GIMIC}
\label{sec:relativeimportance}

\begin{figure}
\includegraphics[width=\columnwidth]{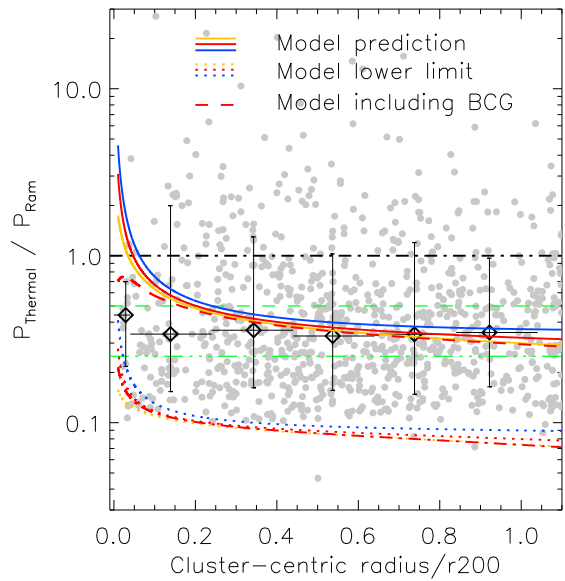}
\caption{Ratio $\alpha$ of confinement (thermal) pressure to ram pressure for our sample of galaxies from the \textsc{gimic} simulation, plotted versus the galaxies' cluster-centric distance in units of $r_{200}$ (grey circles). The black diamonds and error bars show the resulting medians and 25$^{\text{th}}$/75$^\text{th}$ percentiles in bins of width $\Delta (r/r_{200}) = 0.2$ For most galaxies --- but not all --- ram pressure dominates over confinement pressure. The solid yellow, red and blue curves show the expected relation from a simple analytic model of gas in hydrostatic equilibrium inside an NFW halo as described in the text. The corresponding dotted curves indicate the expected lower limits.  The dashed red curve corresponds to the expected relation from a simple analytic model of gas in hydrostatic equilibrium inside an NFW halo that takes into account the mass distribution of the BCG (with the corresponding lower limit shown by the red dash-dot curve).  The horizontal green dashed and dash-dot-dot-dot lines indicate a ratio $\alpha$ of 0.5 and 0.25, as expected from a power-law density profile with exponent -2 and -3 respectively.  Overall, the galaxies from the \textsc{gimic} simulation follow the predictions of the analytic models quite well.} 
\label{fig:prat_radius}
\end{figure}

Fig.~\ref{fig:prat_radius} shows the pressure ratio $\alpha$ obtained as described in Section \ref{sec:simulations} as a function of the cluster-centric radius $r$ of each galaxy, normalised to the virial radius $r_{200}$ of the host. Each of the filled grey circles represents a simulated galaxy, while the black open diamonds and error bars give the median and 25$^\text{th}$/75$^\text{th}$ percentile of the distribution within bins of width $\Delta (r/r_{200}) = 0.2$.

For a clear majority of galaxies, the ratio is less than unity, and so ram pressure is dominant. But despite this general result, there are \emph{individual} galaxies for whom confinement pressure exceeds ram pressure (16 \% of the whole sample at $z = 0$), in some cases by more than a factor of ten.

For comparison, we construct somewhat more realistic analytic cluster models than considered in Section \ref{sec:analyticexpectations}. In particular, we use the methodology of \citet{McCarthy_et_al_2008b} to generate clusters in hydrostatic equilibrium within an NFW potential well for several different central entropy levels: 0 (yellow curves), 30 keV cm$^2$ (red curves) and 100 keV cm$^2$ (blue curves). These roughly span the range of observed central entropies of local galaxy clusters (e.g., \citealt{Cavagnolo_et_al_2009}).  The solid colour curves correspond to galaxies orbiting with the typical velocity of $v_{\rm circ}(r)$.  The dotted colour curves correspond to case where galaxies orbit with a velocity of $v_{\rm circ}(r)+\sigma_{\rm 3D}(r)$, where $\sigma_{\rm 3D}(r)$ is the 3D velocity dispersion profile of the NFW cluster.  This case represents an effective lower limit\footnote{Note that since $\sigma_{\rm 3D}(r)$ is generally larger than $v_{\rm circ}(r)$ we cannot compute an analogous upper bound curve.  Physically, the upper bound occurs at apocentre, when the galaxy's orbital velocity is at a minimum.  For purely radial orbits, the ratio of confinement pressure to ram pressure goes to infinity at apocentre.} on the ratio of the confinement pressure to the ram pressure.  

There is remarkably good agreement in the typical ratio seen in the simulations with what is predicted by the analytic models at all radii except for the very smallest. In both the simulations and the analytic models, the pressure ratio is largely independent of position within the cluster, except for small radii ($r/r_{200} < 0.2$), where the predicted pressure ratio from our model increases sharply and tends to infinity as $r \rightarrow 0$, as expected (see Section 3.1). The galaxies in \textsc{gimic}, however, show only a small increase.

A possible reason for this discrepancy at small radii is the presence of the BCG in \textsc{gimic}. Its mass acts to increase the slope of the potential well in the cluster centre and therefore, according to the logic in Section \ref{sec:analyticexpectations}, to \emph{reduce} the pressure ratio $\alpha$ in the central region. To investigate the extent of this effect, we have modified the 30 keV cm$^2$ model (red curve in Fig.~\ref{fig:prat_radius}) to take into account the presence of the BCG of the massive cluster at the centre of the $+2 \sigma$ simulation (i.e., we recompute the hydrostatic configuration of the gas using the sum of the dark matter NFW and BCG mass distributions).  The pressure ratio resulting from this modified model is shown as a red dashed curve in Fig.~\ref{fig:prat_radius} and is in much better agreement with the simulation results.  In any case, only a small fraction of galaxies spend time there and those that do venture into the very centre will likely be disrupted by tidal forces. 

A comparison of the three analytic model curves also confirms our expectation that the results are quite insensitive to the entropy (or density) distribution of the ICM.

\begin{figure}
\includegraphics[width=\columnwidth]{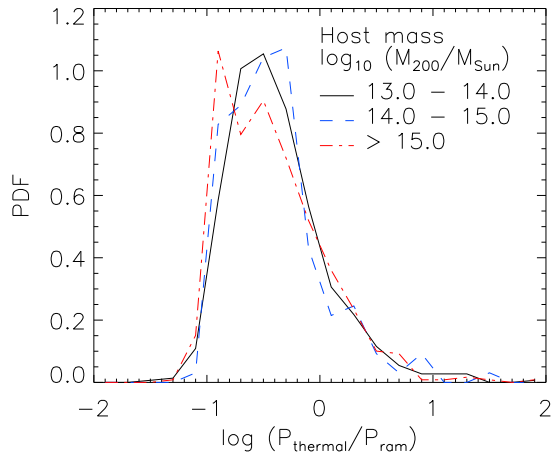}
\caption{Variation of confinement to ram pressure ratio with halo mass. The three curves show the distribution functions for three halo mass ranges as indicated in the figure.  The distribution of confinement to ram pressure ratio does not depend on halo mass.}
\label{fig:prat_histograms}
\end{figure}

The extent of variation of the ratio between confinement and ram pressure with halo mass is shown in Fig.~\ref{fig:prat_histograms}. No systematic variation is evident, which is consistent with our expectations based on the self-similarity of dark matter halos. 

\subsection{Effect of confinement domination}
\label{sec:confinementdomination}
Despite confirming the overall dominance of ram pressure stripping in groups and clusters, the above results show that there is a small fraction of galaxies in which confinement pressure is expected to be more important than ram pressure. However, this does not automatically imply that these galaxies contain more hot gas than ram-pressure dominated ones (or, more relevant in the context of \citetalias{Mulchaey_Jeltema_2010}, field galaxies of the same stellar mass) --- the relative importance of both may well have changed since the galaxy joined the group or cluster. \citet{McCarthy_et_al_2008} found that ram pressure stripping of hot gas halos takes place mostly during the first $\sim$ 2-3 Gyr after infall: At this point, the galaxy has typically passed the first pericentre of its orbit and lost all hot gas that can be stripped. In order to enhance the hot gas content of a galaxy, confinement pressure must therefore be dominant within this period, or there will not be much gas left to be confined. To determine whether confinement pressure will indeed act to retain hot gas haloes in group and cluster galaxies, we therefore need to track the orbital history of the galaxies from the time of infall to the present day.

\begin{figure}
\includegraphics[width=\columnwidth]{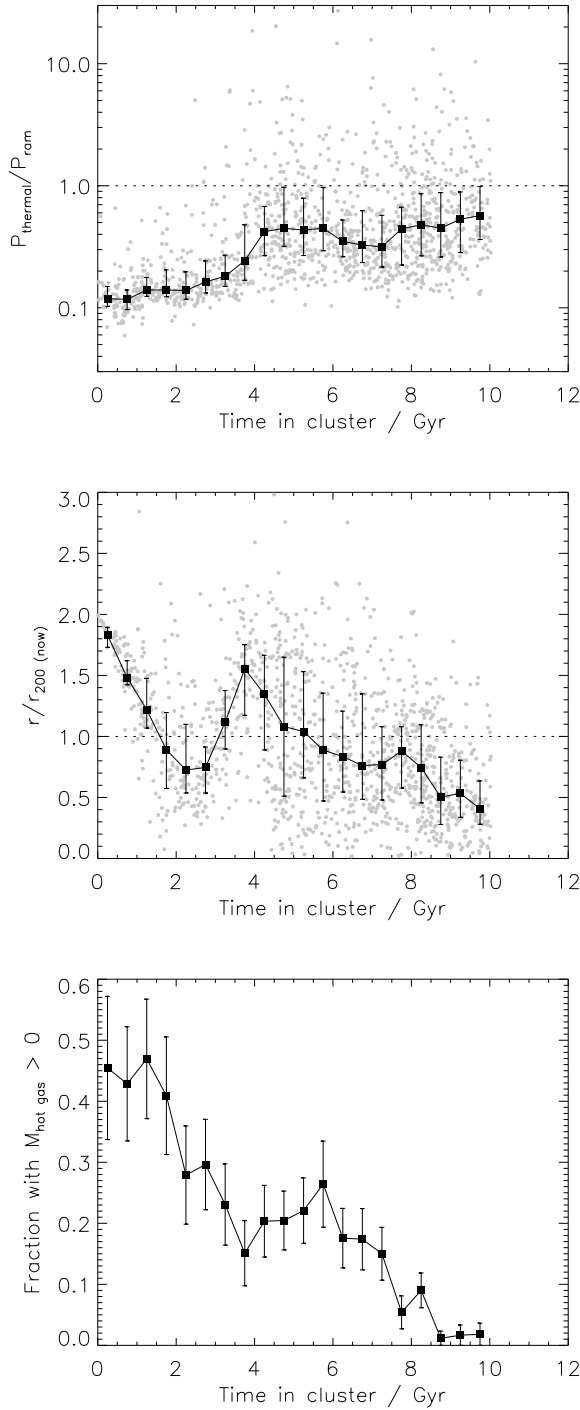}
\caption{The ratio $\alpha = \pth/\pram$ (top panel), cluster-centric distance (middle panel), and fraction of galaxies with some hot gas still bound (bottom panel) as a function of time since infall into the group or cluster, see text for details. Each filled grey circle represents a simulated galaxy.  The filled black squares with error bars represent the median and 25$^\text{th}$/75$^\text{th}$ percentile of the distribution within each bin. Galaxies that have been in the cluster for less than $\sim$ 2-3 Gyr (corresponding to the time between infall and first pericentric passage) are all ram-pressure dominated.  Pressure confinement becomes increasingly important near apocentre (roughly $4$ and $8$ Gyr after infall). The bottom panel shows the fraction of galaxies with some hot gas still bound, which decreases rapidly between 2 and 4 Gyr after infall. The error bars in this panel represent the statistical Poisson uncertainty.  Confinement pressure is generally ineffective.}
\label{fig:prat_time}
\end{figure}

For this purpose, we define `infall' as the first time a galaxy crosses the radius $r = 2\, r_{200} (z)$, where we have adopted this radius, rather than e.g.~$r_{200}$, because many galaxies return to cluster-centric distances $r_{200} < r < 2\,r_{200}$ after pericentric passage (see, e.g., Fig.~\ref{fig:prat_time}) and therefore enter the region within $r_{200}$ more than once. To determine the infall time of the galaxies identified at redshift $z = 0$ we trace their dark matter halos back in time by a method similar to that used by \citet{Font_et_al_2011}. Firstly, we identify for each `galaxy' subhalo at redshift $z = 0$ the dark matter particles associated with it by \textsc{subfind}, and then use their unique simulation IDs to find the subhalo to which each of these dark matter particles belonged to at redshift $z = 2$, discarding unbound particles\footnote{In our adopted cosmology, a redshift of $z = 2$ corresponds to a lookback time of $\sim$ 10 Gyr.}. The subhalo that contains most of these DM particles (excluding the BCG) is identified as the main galaxy progenitor. At each intermediate redshift $z$, we then find the subhalo containing most of the DM particles present in both our original halo (at $z = 0$) and its progenitor at $z = 2$ to trace the galaxy forward in time. The same procedure is applied to the entire FoF group itself, starting with all DM particles with $r_i \leq r_{200}$ at $z = 0$ to find the accurate $r_{200} (z)$ in each snapshot. 

The infall time $t_\text{in}$ is then found by linear interpolation between its position at the last snapshot with $r_\text{gal} > 2\,r_{200}$ and the one immediately after this. Although this method cannot identify the infall time for galaxies already in the cluster at $z = 2$, this only affects a small fraction (6\%) of the galaxies in orbit about the groups and clusters at $z=0$ and is insignificant for our conclusions below.

The top and middle panels of Fig.~\ref{fig:prat_time} show, respectively, the ratio $\alpha$ between confinement and ram pressure and the cluster-centric distance normalised to the present-day virial radius of the host group/cluster, in both cases as a function of time since infall.  Both show a clear trend with time: the distance decreases for the first $\sim$ 2 Gyr after infall as galaxies approach the first pericentre passage, and increases again until $\sim$ 4 Gyr as the galaxies get closer to their first apocentre. Correspondingly, the ratio $\alpha$ is very low (i.e., ram pressure is strongly dominant) for the first $\sim$ 2 Gyr with no single galaxy being confinement-dominated and a median value $\alpha \sim 0.1$. Only around 4 Gyr after infall, at the time of first apocentric passage and the associated drop in galaxy velocity is confinement becoming increasingly more important.  However, by this point in time many of the galaxies have already been completely stripped of their hot gas (bottom panel; see also \citealt{McCarthy_et_al_2008}).  The most active period of ram pressure stripping occurs near first pericentric passage and since first pericentre necessarily precedes first apocentre there will be much less (if any) hot gas to be confined at first apocentre.

A close inspection of the bottom panel of Fig.~\ref{fig:prat_time} shows that even at infall only $\sim 45\%$ of simulated galaxies with $\log_{10} (M_*/M_\odot) \geq 9.0$ have hot gas atmospheres. This relatively low fraction is due to two effects: mass selection and the influence of the group/cluster beyond $2\, r_{200}$.  In terms of mass selection, in the simulations a galaxy of stellar mass $\log_{10} (M_*/M_\odot) = 9.0$ has a total virial mass of $\log_{10} (M_{200}/M_\odot) \sim 11.0 \pm 0.2$, which lies approximately on the halo mass threshold required to support a hot gas atmosphere (e.g., \citealt{Birnboim_Dekel_2003}). Therefore, even amongst the \textsc{gimic} field galaxies of this mass, $\sim 25\%$ have no hot gas haloes, whereas these are found in virtually all more massive field galaxies ($\log_{10} (M_*/M_\odot) \geq 10.0$). The second and more imprtant factor is that ram pressure stripping is effective in the simulations at radii well exceeding $2\, r_{200}$, the radius whose crossing we define as `infall' in this study.  As we will show in a forthcoming paper (Bah\'{e} et al., in prep.), the hot gas atmospheres of massive clusters induce stripping out as far as $\sim 4\, r_{200}$ from the cluster centre. As a result, roughly half of the originally gas-rich galaxies reach $2\, r_{200}$ without any remaining hot gas. Interestingly, there is mounting observational evidence that the effect of environment on galaxies does indeed extend to several virial radii (e.g., \citealt{Haines_et_al_2009}; \citealt{Lu_et_al_2012}).

\begin{figure}
\includegraphics[width=\columnwidth]{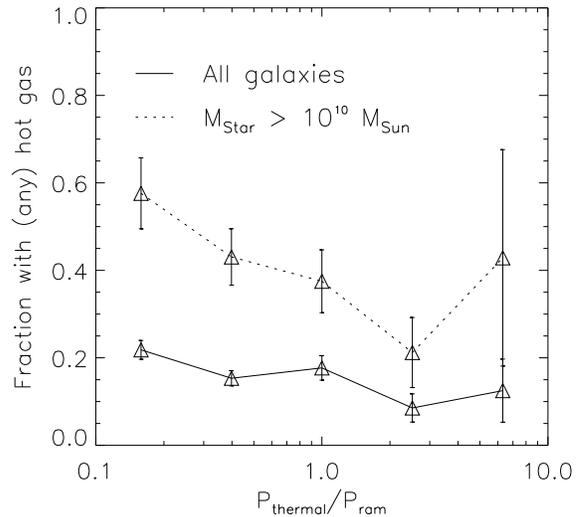}
\caption{Fraction of galaxies with bound hot gas, plotted against the ratio $\alpha = \pth/\pram$. The solid and dashed curves represent the full galaxy sample and those with $M_* > 10^{10} M_\odot$, respectively. While the latter are, in general, more gas-rich, both samples show a clear trend to less hot gas with increasing importance of confinement pressure.} 
\label{fig:prat_hotgas}
\end{figure}

As further confirmation of the general ineffectiveness of confinement pressure, we show in Fig.~\ref{fig:prat_hotgas} the fraction of galaxies containing \emph{any} bound hot gas ($T > 10^5$ K) as a function of the ratio between confinement and ram pressure. When ram pressure is most strongly dominating, this is the case for $\sim 20\%$ of galaxies. With increasing importance of thermal confinement pressure, however, this fraction actually \emph{decreases}, reaching a level of only $\sim 10\%$ out of those galaxies with $\pth > \pram$.  This trend is even stronger for more massive galaxies ($M_* > 10^{10} M_\odot$): within these, hot gas is present in $\sim 60\%$ at $\pth/\pram \sim 0.1$, but only in $\sim 20\%$ of confinement dominated cases.  Therefore, confinement pressure is not only generally \emph{inefficient} compared to ram pressure, but actually \emph{appears counter-effective} in keeping hot gas within the halo in those cases where it is dominant. This is due simply to the fact that galaxies that are currently pressure confined have already been heavily stripped (as shown above they have gone through first pericentre), whereas galaxies that are ram pressure dominated may not yet have been fully stripped if they have fallen in only recently.

\section{Discussion and Conclusions}
\label{sec:observations}

We have investigated the relative importance of ram pressure stripping and thermal pressure confinement on galaxies in groups and clusters at redshift $z = 0$ in the \textsc{gimic} simulation. A large sample of 69 groups and clusters with masses in the range $13.0 \leq \log_{10} (M_{200}/M_\odot) < 15.1$, containing over 1000 galaxies in total, was analysed for this purpose. Our findings may be summarised as follows:

\begin{itemize}
\item Thermal confinement pressure only dominates ram pressure in a small fraction (16\%) of galaxies. In the majority of cases, the action of the ICM is a removal of hot gas from group and cluster galaxies.
\item The ratio between confinement and ram pressure obtained from \textsc{gimic} agrees well with results from a simple analytic model of the ICM in which gas is in hydrostatic equilibrium within an NFW potential well and galaxies orbit at the typical velocity of $v_{\rm circ}(r)$. Outside the very central cluster region, the median value of the ratio is $\alpha \sim 0.3$.
\item Increased confinement pressure does \emph{not} lead to increased retention of hot gas. On the contrary, the fraction of galaxies containing any hot gas decreases with increasing importance of confinement pressure (from $\sim 60\%$ to $\sim 20\%$ for galaxies with $M_* > 10^{10} M_\odot$).
\item The ineffectiveness of confinement pressure to retain hot gas is explained by the orbital history of the galaxies: upon falling into the cluster, they first experience the maximum ram pressure stripping influence around pericentric passage, and only later --- when they reach apocentre --- a dominant influence of confinement pressure. By this point, there is often no hot gas left to be confined.    
\end{itemize}

\begin{figure}
\includegraphics[width=\columnwidth]{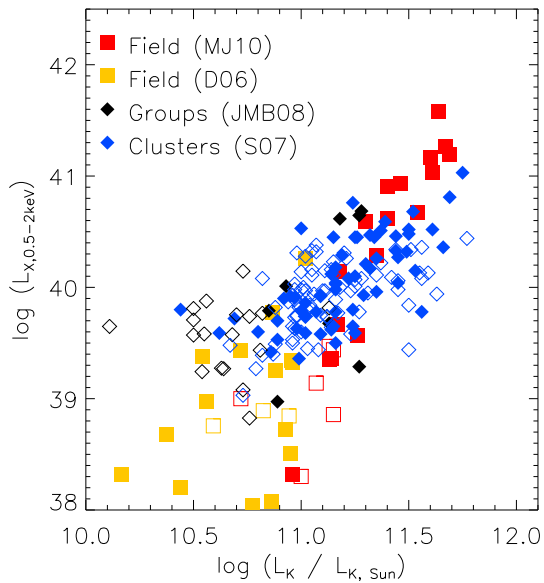}
\caption{Thermal X-ray fluxes (filled symbols) and upper limits (open symbols) as function of galaxy K-band luminosity from the literature, see text for details. Considering the full sample of galaxies shown here, low-$L_K$ field galaxies do not appear systematically less X-ray luminous than their group/cluster counterparts.}
\label{fig:lxlk}
\end{figure}

We point out that the ratio $\alpha$ between confinement and ram pressure depends only on the galaxy velocity $v$ and the ICM temperature $T$.  Both of these quantities are set by the depth of the cluster potential well, which is dominated by dark matter.  Therefore, we expect our results to be insensitive to uncertain sub-grid processes.  Furthermore, since the dark matter mass distribution is approximately self-similar, our results are approximately independent of total host mass, which we explicitly verified in Section \ref{sec:relativeimportance}.  

Our results are therefore seemingly at odds with the observational evidence presented by \citetalias{Mulchaey_Jeltema_2010}, which apparently show that galaxies of fixed stellar mass in groups and clusters are more X-ray luminous than their field counterparts.  A potential explanation for this is shown in Fig.~\ref{fig:lxlk}, an expanded version of Fig.~1 from \citetalias{Mulchaey_Jeltema_2010}, which shows the $L_K - L_X$ relation for field and group/cluster galaxies. The former are taken from \citetalias{Mulchaey_Jeltema_2010}, as well as \citet[D06]{David_et_al_2006}, while we use the catalogues published by \citet[JMB08]{Jeltema_et_al_2008} and \citet[S07]{Sun_et_al_2007} for group and cluster galaxies respectively. All these catalogues contain in part galaxies in which a thermal X-ray component was not detected, and for which therefore only upper limits on the thermal X-ray flux are available (shown by open symbols in Fig.~\ref{fig:lxlk}.  We see no significant difference between the X-ray properties of the \emph{detected} galaxies in the field, group and cluster samples. 

\begin{figure}
\includegraphics[width=\columnwidth]{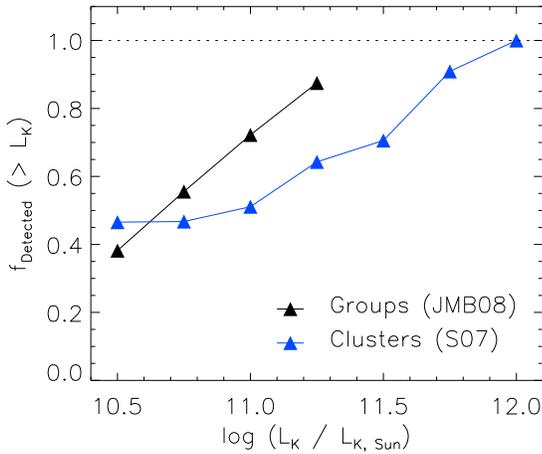}
\caption{X-ray detection fractions for the observational samples shown in Fig.~\ref{fig:lxlk}. Shown is the fraction of galaxies above a given K-band luminosity $L_K$ with detected thermal X-ray emission. While galaxies with larger $L_K$ are, in all environments, more likely to be X-ray detected, there is a clear trend to lower detection fractions in denser environments at fixed threshold $L_K$. This confirms that galaxies in groups and clusters are hot gas poor compared to the field population, as expected from ram pressure stripping.}
\label{fig:detection_fractions}
\end{figure}

However, in Fig.~\ref{fig:detection_fractions} we show the detection fractions of group and cluster glaxies from the samples of \citetalias{Jeltema_et_al_2008} and \citetalias{Sun_et_al_2007}. In both samples galaxies with higher $L_K$ --- and therefore deeper potential wells --- are more likely to be detected in X-rays. Galaxies of a \emph{given} stellar mass (as indicated by their K-band luminosity), on the other hand, are more likely to be detected in groups than in clusters.  Both of these trends are what one qualitatively expects from ram pressure stripping: high mass galaxies are able to retain more gas (which accounts for the increased detection fraction with increasing $L_K$) and galaxies of fixed mass are more effectively stripped in clusters than in groups, due primarily to the higher orbital velocities of galaxies in clusters.
We note, however, that the comparison between the group and cluster galaxy samples is made complicated by the fact that the surface brightness of the background hot gas (the ICM) is generally higher in more massive systems.  Defining the detection fraction to be the fraction of galaxies above a certain {\it fixed} surface brightness threshold (which could be that of the surface brightness of the most massive cluster in the sample) would be one way to rectify this problem.

There is also an issue regarding the comparison of the group and cluster galaxy samples to the field galaxy sample.  In particular, the group and cluster galaxy samples are based on an optical selection, whereas the field galaxy samples are drawn from a mixture of previous optical and X-ray catalogs.  X-ray follow-up of an optically-selected field sample is required to make a proper like-with-like comparison with the group and cluster results of \citetalias{Jeltema_et_al_2008} and \citetalias{Sun_et_al_2007}.

Our theoretical results strongly suggest that ram pressure stripping is generally dominant over pressure confinement and that strangulation should be effective in the group/cluster environment.  This bodes well for semi-analytic models of galaxy formation that invoke strangulation to explain the environmental dependence of galaxy colours for galaxies of fixed stellar mass.  In a future study, we will examine the effect of strangulation on simulated galaxies in \textsc{gimic}.

\section*{acknowledgements}
The authors thank Ming Sun for helpful comments and the anonymous referee for a constructive report.
YMB acknowledges a postgraduate award from STFC. IGM is supported by an Advanced Fellowship from the STFC and a Birmingham Fellowship at the University of Birmingham. This research has made use of the \textsc{Darwin} High Performance Computing Facility at the University of Cambridge. The \textsc{gimic} simulations were carried out using the HPCx facility at the Edinburgh Parallel Computing Centre (EPCC) as part of the EC's DEISA `Extreme Computing Initiative' and with the Cosmology Machine at the Institute of Computational Cosmology of Durham University.

\bibliographystyle{mn2e}
\bibliography{Confinement}

\begin{thebibliography}{49}
\expandafter\ifx\csname natexlab\endcsname\relax\def\natexlab#1{#1}\fi

\bibitem[{{Abadi}, {Moore} \& {Bower}(1999){Abadi}, {Moore}, \&
  {Bower}}]{Abadi_et_al_1999}
{Abadi} M.~G., {Moore} B., {Bower} R.~G., 1999, \mnras, 308, 947

\bibitem[{{Anderson} \& {Bregman}(2011)}]{Anderson_Bregman_2011}
{Anderson} M.~E., {Bregman} J.~N., 2011, \apj, 737, 22

\bibitem[{{Balogh}, {Navarro} \& {Morris}(2000){Balogh}, {Navarro}, \&
  {Morris}}]{Balogh_et_al_2000}
{Balogh} M.~L., {Navarro} J.~F., {Morris} S.~L., 2000, \apj, 540, 113

\bibitem[{{Birnboim} \& {Dekel}(2003)}]{Birnboim_Dekel_2003}
{Birnboim} Y., {Dekel} A., 2003, \mnras, 345, 349

\bibitem[{{Cavagnolo} {et~al}\mbox{.}(2009){Cavagnolo}, {Donahue}, {Voit}, \&
  {Sun}}]{Cavagnolo_et_al_2009}
{Cavagnolo} K.~W., {Donahue} M., {Voit} G.~M., {Sun} M., 2009, \apjs, 182, 12

\bibitem[{{Crain} {et~al}\mbox{.}(2010{\natexlab{a}}){Crain}, {McCarthy},
  {Frenk}, {Theuns}, \& {Schaye}}]{Crain_et_al_2010a}
{Crain} R.~A., {McCarthy} I.~G., {Frenk} C.~S., {Theuns} T., {Schaye} J.,
  2010{\natexlab{a}}, \mnras, 407, 1403

\bibitem[{{Crain} {et~al}\mbox{.}(2010{\natexlab{b}}){Crain}, {McCarthy},
  {Schaye}, {Frenk}, \& {Theuns}}]{Crain_et_al_2010b}
{Crain} R.~A., {McCarthy} I.~G., {Schaye} J., {Frenk} C.~S., {Theuns} T.,
  2010{\natexlab{b}}, ArXiv e-prints

\bibitem[{{Crain} {et~al}\mbox{.}(2009){Crain}, {Theuns}, {Dalla Vecchia},
  {Eke}, {Frenk}, {Jenkins}, {Kay}, {Peacock}, {Pearce}, {Schaye}, {Springel},
  {Thomas}, {White}, \& {Wiersma}}]{Crain_et_al_2009}
{Crain} R.~A. {et~al.}, 2009, \mnras, 399, 1773

\bibitem[{{Dalla Vecchia} \& {Schaye}(2008)}]{DallaVecchia_Schaye_2008}
{Dalla Vecchia} C., {Schaye} J., 2008, \mnras, 387, 1431

\bibitem[{{David} {et~al}\mbox{.}(2006){David}, {Jones}, {Forman}, {Vargas}, \&
  {Nulsen}}]{David_et_al_2006}
{David} L.~P., {Jones} C., {Forman} W., {Vargas} I.~M., {Nulsen} P., 2006,
  \apj, 653, 207

\bibitem[{{Dolag} {et~al}\mbox{.}(2009){Dolag}, {Borgani}, {Murante}, \&
  {Springel}}]{Dolag_et_al_2009}
{Dolag} K., {Borgani} S., {Murante} G., {Springel} V., 2009, \mnras, 399, 497

\bibitem[{{Font} {et~al}\mbox{.}(2008){Font}, {Bower}, {McCarthy}, {Benson},
  {Frenk}, {Helly}, {Lacey}, {Baugh}, \& {Cole}}]{Font_et_al_2008}
{Font} A.~S. {et~al.}, 2008, \mnras, 389, 1619

\bibitem[{{Font} {et~al}\mbox{.}(2011){Font}, {McCarthy}, {Crain}, {Theuns},
  {Schaye}, {Wiersma}, \& {Dalla Vecchia}}]{Font_et_al_2011}
{Font} A.~S., {McCarthy} I.~G., {Crain} R.~A., {Theuns} T., {Schaye} J.,
  {Wiersma} R.~P.~C., {Dalla Vecchia} C., 2011, \mnras, 416, 2802

\bibitem[{{Forman}, {Jones} \& {Tucker}(1985){Forman}, {Jones}, \&
  {Tucker}}]{Forman_et_al_1985}
{Forman} W., {Jones} C., {Tucker} W., 1985, \apj, 293, 102

\bibitem[{{Gunn} \& {Gott}(1972)}]{Gunn_Gott_1972}
{Gunn} J.~E., {Gott}, III J.~R., 1972, \apj, 176, 1

\bibitem[{{Haines} {et~al}\mbox{.}(2009){Haines}, {Smith}, {Egami}, {Ellis},
  {Moran}, {Sanderson}, {Merluzzi}, {Busarello}, \&
  {Smith}}]{Haines_et_al_2009}
{Haines} C.~P. {et~al.}, 2009, \apj, 704, 126

\bibitem[{{Hansen} {et~al}\mbox{.}(2011){Hansen}, {Macci{\'o}}, {Romano-Diaz},
  {Hoffman}, {Br{\"u}ggen}, {Scannapieco}, \& {Stinson}}]{Hansen_et_al_2011}
{Hansen} S.~H., {Macci{\'o}} A.~V., {Romano-Diaz} E., {Hoffman} Y.,
  {Br{\"u}ggen} M., {Scannapieco} E., {Stinson} G.~S., 2011, \apj, 734, 62

\bibitem[{{Jeltema}, {Binder} \& {Mulchaey}(2008){Jeltema}, {Binder}, \&
  {Mulchaey}}]{Jeltema_et_al_2008}
{Jeltema} T.~E., {Binder} B., {Mulchaey} J.~S., 2008, \apj, 679, 1162

\bibitem[{{Kim}, {Fabbiano} \& {Trinchieri}(1992){Kim}, {Fabbiano}, \&
  {Trinchieri}}]{Kim_et_al_1992}
{Kim} D.-W., {Fabbiano} G., {Trinchieri} G., 1992, \apj, 393, 134

\bibitem[{{Larson}, {Tinsley} \& {Caldwell}(1980){Larson}, {Tinsley}, \&
  {Caldwell}}]{Larson_et_al_1980}
{Larson} R.~B., {Tinsley} B.~M., {Caldwell} C.~N., 1980, \apj, 237, 692

\bibitem[{{Li} \& {Wang}(2012)}]{Li_Wang_2012}
{Li} J.-T., {Wang} Q.~D., 2012, submitted

\bibitem[{{Li}, {Wang} \& {Hameed}(2007){Li}, {Wang}, \&
  {Hameed}}]{Li_et_al_2007}
{Li} Z., {Wang} Q.~D., {Hameed} S., 2007, \mnras, 376, 960

\bibitem[{{Li} {et~al}\mbox{.}(2006){Li}, {Wang}, {Irwin}, \&
  {Chaves}}]{Li_et_al_2006}
{Li} Z., {Wang} Q.~D., {Irwin} J.~A., {Chaves} T., 2006, \mnras, 371, 147

\bibitem[{{Lu} {et~al}\mbox{.}(2012){Lu}, {Gilbank}, {McGee}, {Balogh}, \&
  {Gallagher}}]{Lu_et_al_2012}
{Lu} T., {Gilbank} D.~G., {McGee} S.~L., {Balogh} M.~L., {Gallagher} S., 2012,
  \mnras, 420, 126

\bibitem[{{Mathews} \& {Brighenti}(1998)}]{Mathews_Brighenti_1998}
{Mathews} W.~G., {Brighenti} F., 1998, \apjl, 503, L15+

\bibitem[{{McCarthy} {et~al}\mbox{.}(2008{\natexlab{a}}){McCarthy}, {Babul},
  {Bower}, \& {Balogh}}]{McCarthy_et_al_2008b}
{McCarthy} I.~G., {Babul} A., {Bower} R.~G., {Balogh} M.~L.,
  2008{\natexlab{a}}, \mnras, 386, 1309

\bibitem[{{McCarthy} {et~al}\mbox{.}(2012){McCarthy}, {Font}, {Crain},
  {Deason}, {Schaye}, \& {Theuns}}]{McCarthy_et_al_2012}
{McCarthy} I.~G., {Font} A.~S., {Crain} R.~A., {Deason} A.~J., {Schaye} J.,
  {Theuns} T., 2012, \mnras, 420, 2245

\bibitem[{{McCarthy} {et~al}\mbox{.}(2008{\natexlab{b}}){McCarthy}, {Frenk},
  {Font}, {Lacey}, {Bower}, {Mitchell}, {Balogh}, \&
  {Theuns}}]{McCarthy_et_al_2008}
{McCarthy} I.~G., {Frenk} C.~S., {Font} A.~S., {Lacey} C.~G., {Bower} R.~G.,
  {Mitchell} N.~L., {Balogh} M.~L., {Theuns} T., 2008{\natexlab{b}}, \mnras,
  383, 593

\bibitem[{{Moore} {et~al}\mbox{.}(1996){Moore}, {Katz}, {Lake}, {Dressler}, \&
  {Oemler}}]{Moore_et_al_1996}
{Moore} B., {Katz} N., {Lake} G., {Dressler} A., {Oemler} A., 1996, \nat, 379,
  613

\bibitem[{{Mulchaey} \& {Jeltema}(2010)}]{Mulchaey_Jeltema_2010}
{Mulchaey} J.~S., {Jeltema} T.~E., 2010, \apjl, 715, L1

\bibitem[{{O'Sullivan}, {Forbes} \& {Ponman}(2001){O'Sullivan}, {Forbes}, \&
  {Ponman}}]{Osullivan_et_al_2001}
{O'Sullivan} E., {Forbes} D.~A., {Ponman} T.~J., 2001, \mnras, 328, 461

\bibitem[{{Owen} \& {Warwick}(2009)}]{Owen_Warwick_2009}
{Owen} R.~A., {Warwick} R.~S., 2009, \mnras, 394, 1741

\bibitem[{{Read} \& {Ponman}(1998)}]{Read_Ponman_1998}
{Read} A.~M., {Ponman} T.~J., 1998, \mnras, 297, 143

\bibitem[{{Schaye} \& {Dalla Vecchia}(2008)}]{Schaye_DallaVecchia_2008}
{Schaye} J., {Dalla Vecchia} C., 2008, \mnras, 383, 1210

\bibitem[{{Schaye} {et~al}\mbox{.}(2010){Schaye}, {Dalla Vecchia}, {Booth},
  {Wiersma}, {Theuns}, {Haas}, {Bertone}, {Duffy}, {McCarthy}, \& {van de
  Voort}}]{Schaye_et_al_2010}
{Schaye} J. {et~al.}, 2010, \mnras, 402, 1536

\bibitem[{{Springel}(2005)}]{Springel_2005}
{Springel} V., 2005, \mnras, 364, 1105

\bibitem[{{Springel} {et~al}\mbox{.}(2005){Springel}, {White}, {Jenkins},
  {Frenk}, {Yoshida}, {Gao}, {Navarro}, {Thacker}, {Croton}, {Helly},
  {Peacock}, {Cole}, {Thomas}, {Couchman}, {Evrard}, {Colberg}, \&
  {Pearce}}]{Springel_et_al_2005}
{Springel} V. {et~al.}, 2005, Nature, 435, 629

\bibitem[{{Springel} {et~al}\mbox{.}(2001){Springel}, {White}, {Tormen}, \&
  {Kauffmann}}]{Springel_et_al_2001}
{Springel} V., {White} S.~D.~M., {Tormen} G., {Kauffmann} G., 2001, \mnras,
  328, 726

\bibitem[{{Strickland} {et~al}\mbox{.}(2004){Strickland}, {Heckman}, {Colbert},
  {Hoopes}, \& {Weaver}}]{Strickland_et_al_2004}
{Strickland} D.~K., {Heckman} T.~M., {Colbert} E.~J.~M., {Hoopes} C.~G.,
  {Weaver} K.~A., 2004, \apjs, 151, 193

\bibitem[{{Sun} {et~al}\mbox{.}(2007){Sun}, {Jones}, {Forman}, {Vikhlinin},
  {Donahue}, \& {Voit}}]{Sun_et_al_2007}
{Sun} M., {Jones} C., {Forman} W., {Vikhlinin} A., {Donahue} M., {Voit} M.,
  2007, \apj, 657, 197

\bibitem[{{Sun} {et~al}\mbox{.}(2009){Sun}, {Voit}, {Donahue}, {Jones},
  {Forman}, \& {Vikhlinin}}]{Sun_et_al_2009}
{Sun} M., {Voit} G.~M., {Donahue} M., {Jones} C., {Forman} W., {Vikhlinin} A.,
  2009, \apj, 693, 1142

\bibitem[{{T{\"u}llmann} {et~al}\mbox{.}(2006){T{\"u}llmann}, {Pietsch},
  {Rossa}, {Breitschwerdt}, \& {Dettmar}}]{Tuellmann_et_al_2006}
{T{\"u}llmann} R., {Pietsch} W., {Rossa} J., {Breitschwerdt} D., {Dettmar}
  R.-J., 2006, \aap, 448, 43

\bibitem[{{Voit} {et~al}\mbox{.}(2002){Voit}, {Bryan}, {Balogh}, \&
  {Bower}}]{Voit_et_al_2002}
{Voit} G.~M., {Bryan} G.~L., {Balogh} M.~L., {Bower} R.~G., 2002, \apj, 576,
  601

\bibitem[{{Wang}(2005)}]{Wang_2005}
{Wang} Q.~D., 2005, in Astronomical Society of the Pacific Conference Series,
  Vol. 331, Extra-Planar Gas, {R.~Braun}, ed., pp. 329--+

\bibitem[{{Weinmann} {et~al}\mbox{.}(2006){Weinmann}, {van den Bosch}, {Yang},
  \& {Mo}}]{Weinmann_et_al_2006}
{Weinmann} S.~M., {van den Bosch} F.~C., {Yang} X., {Mo} H.~J., 2006, \mnras,
  366, 2

\bibitem[{{White} \& {Frenk}(1991)}]{White_Frenk_1991}
{White} S.~D.~M., {Frenk} C.~S., 1991, \apj, 379, 52

\bibitem[{{White} \& {Rees}(1978)}]{White_Rees_1978}
{White} S.~D.~M., {Rees} M.~J., 1978, \mnras, 183, 341

\bibitem[{{Wiersma}, {Schaye} \& {Smith}(2009){Wiersma}, {Schaye}, \&
  {Smith}}]{Wiersma_et_al_2009a}
{Wiersma} R.~P.~C., {Schaye} J., {Smith} B.~D., 2009, \mnras, 393, 99

\bibitem[{{Wiersma} {et~al}\mbox{.}(2009){Wiersma}, {Schaye}, {Theuns}, {Dalla
  Vecchia}, \& {Tornatore}}]{Wiersma_et_al_2009b}
{Wiersma} R.~P.~C., {Schaye} J., {Theuns} T., {Dalla Vecchia} C., {Tornatore}
  L., 2009, \mnras, 399, 574

\end{thebibliography}

\end{document}